\documentclass[journal=jctc,manuscript=article]{achemso}
\setkeys{acs}{articletitle=true}
\usepackage[latin9]{inputenc}
\usepackage[T1]{fontenc}

\usepackage{algorithm}
\usepackage{algorithmic}
\usepackage{amsmath}
\usepackage{amssymb}
\usepackage{bm}
\usepackage{color}
\usepackage{changes}
\usepackage{graphicx}
\usepackage{listings}
\usepackage{verbatim}
\usepackage{appendix}

\makeatletter
\def\x#1{\def\tempa{#1}\futurelet\next\x@i}
\def\x@i{\ifx\next\bgroup\expandafter\x@ii\else\expandafter\x@end\fi}
\def\x@ii#1{x_{\tempa}^{(#1)}}
\def\x@end{x_{\tempa}}
\makeatother

\newcommand{\reals}{\mathbb{R}}
\newcommand{\integers}{\mathbb{Z}}
\newcommand{\xspace}[1]{ {\cal X}_{#1}}

\newcommand{\xdim}{d}

\newcommand{\tens}[1]{\bm{\mathcal{#1}}}

\newcommand{\mat}[1]{\mathbf{#1}}
\newcommand{\bvec}[1]{\mathbf{#1}}
\newcommand{\pdimb}{p}
\newcommand{\pdim}{\pdimb_{t}}
\newcommand{\pdimp}[3]{\pdimb_{#1#2#3}}

\newcommand{\parambase}{\theta}
\newcommand{\param}{\bm{\parambase}}

\newcommand{\paramuni}[4]{\parambase_{#1#2#3#4}}
\newcommand{\mvf}[1]{{\cal #1}}
\newcommand{\ftcore}[1]{{\mvf{F}_{#1}}}

\newcommand{\fcore}[1]{\ftcore{#1}} 

\newcommand{\gcore}[1]{\mvf{G}_{#1}} 
\newcommand{\hcore}[1]{\mvf{H}_{#1}} 
\newcommand{\uni}[4]{{#1}_{#2}^{(#3#4)}}
\newcommand{\unif}[3]{\uni{f}{#1}{#2}{#3}}
\newcommand{\ffiber}[2]{f_{#2}^{(#1)}} 

\newcommand{\multi}{\bm{\alpha}}
\newcommand{\ftexp}{\sum_{i_{0}=1}^{r_0}\sum_{i_{1}=1}^{r_1}\cdots \sum_{i_{d}=1}^{r_d} \unif{1}{i_0}{i_1}(x_1)\unif{2}{i_1}{i_2}(x_2)\ldots\unif{d}{i_{d-1}}{i_d}(x_d)}

\newcommand*{\R}{\ensuremath{\mathbb{R}}}
\newcommand*{\C}{\ensuremath{\mathbb{C}}}

\SectionNumbersOn
\setkeys{acs}{usetitle=true}

\makeatletter
\title{Functional Tensor-Train Chebyshev Method for Multidimensional Quantum Dynamics Simulations}

\author{Micheline B. Soley}
\affiliation{\quad Yale Quantum Institute, Yale University,\\ P.O. Box 208334, New Haven, CT, 06520-8263, USA\\[1mm]}
\alsoaffiliation{\quad Department of Chemistry, Yale University,\\ P.O. Box 208107, New Haven, CT, 06520, USA\\[1mm]}
\author{Paul Bergold}
\affiliation{\quad Zentrum Mathematik, Technical University of Munich,\\ Boltzmannstr. 3, 85748 Garching, Germany\\[1mm]}
\author{Alex A. Gorodetsky}
\affiliation{\quad Department of Aerospace Engineering, University of Michigan,\\ 1320 Beal Avenue Ann Arbor, MI 48109-2140, USA\\[1mm]}
\author{Victor S. Batista}
\affiliation{\quad Yale Quantum Institute, Yale University,\\ P.O. Box 208334, New Haven, CT, 06520-8263, USA\\[1mm]}
\alsoaffiliation{\quad Department of Chemistry, Yale University,\\ P.O. Box 208107, New Haven, CT, 06520, USA\\[1mm]}
\alsoaffiliation{\quad Energy Sciences Institute, Yale University,\quad\\ P.O. Box 27394, West Haven, CT, 06516-7394, USA}
\email{victor.batista@yale.edu}

\makeatother
\begin{document}

\newpage
\begin{abstract}
Methods for efficient simulations of multidimensional quantum dynamics are essential for theoretical studies of chemical systems where quantum effects are important, such as those involving rearrangements of protons or electronic configurations. Here, we introduce the {\em functional tensor-train Chebyshev} (FTTC) method for rigorous nuclear quantum dynamics simulations. FTTC is essentially the Chebyshev propagation scheme applied to the initial state represented in a continuous analogue tensor-train format. We demonstrate the capabilities of FTTC as applied to simulations of proton quantum dynamics in a 50-dimensional model of hydrogen-bonded DNA base pairs.
\end{abstract}

\newpage
\section{Introduction}
Quantum dynamics simulations are essential for rigorous theoretical studies of quantum reaction dynamics, including applications to structural and dynamical problems with critical rearrangements of protons or electronic configurations. A variety of approaches have been developed, including time-dependent Hartree methods\cite{Dirac.1930.376,McLachlan.1964.844,Gerber.1982.3022,Flores.2004.6745,Meyer.1990.73,Meyer.1993.141,Beck.2000.1,Wang.2003.1289,Schulze.2016.185101,Burghardt.1999.2927,Worth.2003.502} and other methods based on short-time approximations of the time-evolution operator, such as the Trotter expansion and finite difference methods.\cite{Kosloff.1994.145,Feit.1982.412,Gray.1994.100,Takahashi.1993.8680,Mazur.1959.1395,McCullough.1969.1253,McCullough.1971.3578,Lanczos.1950.255,Leforestier.1991.59,Kong.2016.3260,Wu.2003.6720,Wu.2004.1676,Chen.2006.124313} Here, we focus on the Chebyshev method for simulation of quantum wavepacket dynamics, which enables computation of the time-evolved quantum state at the final time without having to compute intermediate states at earlier times.\cite{Kosloff.1994.145} Thus, contrary to methods based on short-time propagators, the Chebyshev propagation scheme can be implemented without error accumulation. Chebyshev propagation is currently one of the foremost approaches for simulations of quantum dynamics in low dimensionality,\cite{Kosloff.1994.145,TalEzer.1984.3967,Ndong.2010.064105,Schaefer.2017.368} as demonstrated for nuclear quantum dynamics simulations of molecular systems with up to six dimensions.\cite{Goldfield.2002.1604,Cvitas.2013.064307} However, applications to higher-dimensional systems have been hindered by the exponential scaling of memory and computational cost with dimensionality, due to its reliance on full-grid representations. Here, we introduce a viable solution to the exponential scaling with dimensionality by applying the Chebyshev propagation scheme to the initial state represented in functional tensor-train format (FT) -- i.e., the continuous analogue of the tensor-train/matrix product state decomposition. The resulting functional tensor-train Chebyshev (FTTC) method is demonstrated as applied to simulations of proton dynamics in a high-dimensional (50-dimensional) model of hydrogen-bonded adenine-thymine DNA base pairs, where photo-induced proton transfer has long been thought to have important biological implications ({\em e.g.}, photoinduced mutations).\cite{RevModPhys.35.724}

Tensor networks have been successfully applied to a wide range of studies, including applications to solving partial differential equations,\cite{Khoromskij.2011.257,Townsend.2015.106,Dolgov.2021.A2190,Halimeh.2015.115130} quantum dynamics methods,\cite{Cao1,GevaCao, Cao2,vv11,vv10,Greene.2017.4034,Haegeman.2011.070601,Lubich.2013.470,Lubich.2015.917,Haegeman.2016.165116,Frahm.2019.2154,Cazalilla.2002.256403,Vidal.2003.147902,Luo.2003.049701,White.2004.076401,Vidal.2004.040502,Verstraete.2004.207204,Feiguin.2005.020404R,GarciaRipoll.2006.305,Baiardi.2019.3481,Paeckel.2019.167998} machine learning,\cite{Chertkov.2021.2102.08143v1} electronic structure calculations,\cite{Niklasson.2016.234101,Garnet1,Garnet2,Garnet3,Laura1,Laura2} and calculations of vibrational states.\cite{Tucker1,Tucker2,Tucker3}

Here, we build upon the tensor-train split-operator Fourier transform (TT-SOFT) method,\cite{Greene.2017.4034} and we develop the FTTC method which is essentially a functional tensor-train implementation of the Chebyshev propagation scheme, popularized by Kosloff and co-workers.\cite{TalEzer.1984.3967} FTTC expands the initial state as a functional tensor train and evolves it by applying the Chebyshev expansion of the time-evolution operator, 
\begin{align}\label{eq1}
	e^{-it\hat H}
	\approx\sum_{k=0}^{N-1} \left(2-\delta_{k,0}\right)(-i)^kJ_k(t)T_k(\tens{\hat H})
\end{align}
where $T_k(\tens{\hat H})$ are Chebyshev polynomials of the Hamiltonian $\hat H$ in functional tensor-train format, $J_k(t)$ are the Bessel functions of the first kind, and $t$ is the final propagation time. In practice, a finite number $N\ge 1$ of polynomials is employed and the expansion is applied iteratively in time. Important advantages of the proposed FTTC algorithm when compared to other propagation methods based on matrix product states are: (i) the error need not accumulate with time since the state at time $t$ can be obtained directly without having to compute earlier intermediate states and (ii) the uniform character of the Chebyshev expansion that decreases the error exponentially with $N$.

We focus on functional tensor trains (FT),\cite{Gorodetsky.2019.59,Oseledets.2013.1,Gorodetsky.2017.Continuous,Gorodetsky.2018.1219} which are continuous analogues of tensor trains/matrix product states and have yet to be demonstrated as applied to simulations of quantum nuclear dynamics. Functional tensor-train representations of time-dependent states allow for efficient computations of gradients of multidimensional tensors, so they are expected to be particularly valuable for a variety of applications, including studies of the quantum control of molecular systems.\cite{rego2009coherent} Therefore, we can implement Eq.~\eqref{eq1} as applied to the time-evolving state directly by computing functional tensor-train decompositions of the Chebyshev polynomials applied to the time-evolved wave function without having to pre-compute the Chebyshev polynomials of the Hamiltonian. The FTTC method can also be implemented using discrete tensor trains,\cite{Oseledets.2010.70,Oseledets.2011.2295} instead of their continuous analogues, allowing for efficient representation and manipulation of matrix product states.\cite{Ostlund.1995.3537}

The article is organized as follows. Section~\ref{sec:ChebApproach} describes the Chebyshev methodology. Section~\ref{sub:Chebyshev Polynomials} introduces the Chebyshev polynomials. Section~\ref{sub:Chebyshev Expansion of Complex-Valued Functions} describes how to generate Chebyshev expansions of complex-valued functions, and Section~\ref{sub:Chebyshev Propagation Method} describes Chebyshev propagation based on discrete space representations. Section~\ref{sec:Function-Train} describes the functional tensor-train decomposition, as a continuous analogue tensor-train format for multilinear algebra manipulations of high-dimensional tensors. Section~\ref{sec:TTCheb} describes our functional tensor-train Chebyshev (FTTC) propagation method as implemented for numerical integration of the time-dependent Schr\"odinger equation. Section~\ref{sec:Results} demonstrates the capabilities of FTTC as applied to simulations of proton quantum dynamics in a 50-dimensional model of DNA base pairs with highly anharmonic modes. The Supporting Information demonstrates that tensor-train Chebyshev dynamics can entail lower computational cost relative to the state-of-the-art split operator Fourier transform method for long time steps. The results show that FTTC enables simulations of molecular systems far beyond the capabilities of the standard grid-based Chebyshev method.

\section{Chebyshev Approach\label{sec:ChebApproach}}
The Chebyshev propagation method\cite{TalEzer.1984.3967} integrates numerically the time-dependent Schr\"odinger equation
\begin{equation}\label{eq:schv}
	i\frac{\partial\Psi}{\partial t}
	=\hat{H}\Psi
\end{equation}
where we have used atomic units ($\hbar=1$). For simplicity, we consider a system described by the Hamiltonian
\begin{align}\label{eq:hamiltonian}
	\hat{H}
	=\frac{\hat{p}\cdot\hat{p}}{2m}+\hat V
	=-\frac{1}{2m}\Delta_x\Psi+\hat V
\end{align}
where $m>0$ is the mass of the system, $\hat{p}=-i\nabla_x$ is the momentum operator, and $V\colon\R^d\to\R$ is a given potential energy surface (PES) describing interactions that rule the underlying dynamics of the system. Using the unitary evolution propagator $\hat U(t)=e^{-it\hat H}$, the solution of Eq.~(\ref{eq:schv}), $\Psi(t)$, corresponding to the initial state $\Psi_0$ is
\begin{align}
	\Psi(t)
	=\hat U(t)\Psi_0.
\end{align}
The Chebyshev propagation method approximates the propagator $\hat U(t)= e^{-it\hat H}$ for a fixed final time $t$ in terms of a linear combination of the first $N\ge 1$ Chebyshev polynomials of the Hamiltonian $T_0(\hat H),\dots,T_{N-1}(\hat H)$, as discussed in Section~\ref{sub:Chebyshev Polynomials}.

\subsection{Chebyshev Polynomials}\label{sub:Chebyshev Polynomials}
For all integers $k\ge 0$ and all $x\in [-1,1]$, the $k$th Chebyshev polynomial is defined as follows
\begin{align}
	T_k(x)
	=\cos\big(k\arccos(x)\big)
\end{align}
where $\arccos$ is the inverse of the cosine (\emph{i.e.,} $\cos(\arccos(x))=\arccos(\cos(x))=x$). We note that the Chebyshev polynomials are defined only for input values $x\in[-1,1]$, since the cosine function attains values only in that limited range, and satisfy the following recurrence relation
\begin{align}\label{eq:recurrence}
	T_{k+1}(x)
	=2xT_k(x)-T_{k-1}(x)
\end{align}
with $T_0(x)=1$ and $T_1(x)=x$ defining the subsequent Chebyshev polynomials, so the first four polynomials are
\begin{align}
	T_0(x)
	=1,\quad
	T_1(x)
	=x,\quad
	T_2(x)
	=2x^2-1,\quad
	T_3(x)
	=4x^3-3x.
\end{align}
%
%
\begin{figure}[H]
	\centering
	\hspace{2.5cm}
	\includegraphics{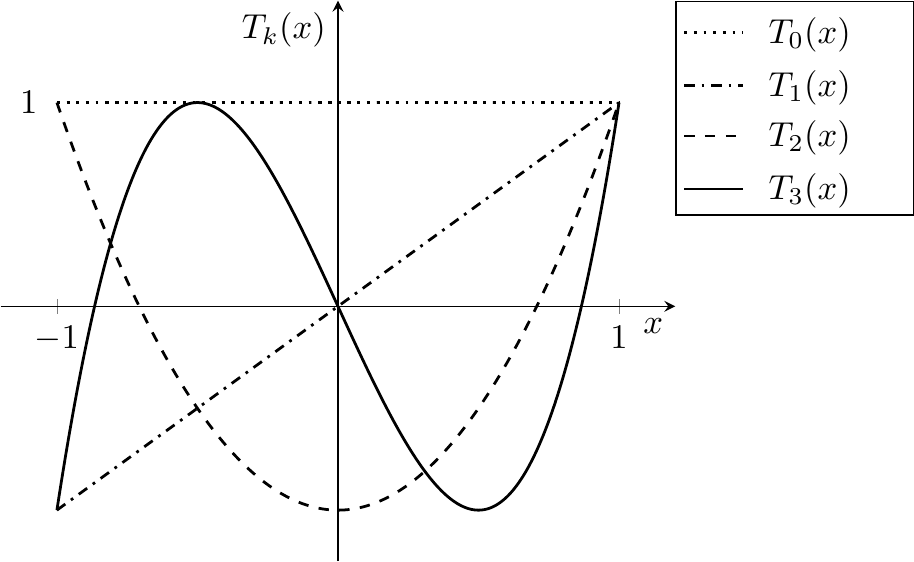}
	\caption{Plot of the first four Chebyshev polynomials, defined only in the limited interval $[-1,1]$, since the possible values of the cosine function are limited to that interval.}
	\label{fig:poly}
\end{figure}
%
We note that Chebyshev polynomials have a number of remarkable properties and are therefore an important tool in the field of approximation theory.\cite{Trefethen:2013uo,Fox:1968up} For instance, let us remark that they satisfy the following orthogonality relation for all $j,k\ge 1,\,j\ne k$,
\begin{align}\label{eq:inner}
	\int_{-1}^1\frac{\mathrm{d}x}{\sqrt{1-x^2}}\,T_j(x)T_k(x)
	=\frac{\pi}{2}\delta_{j,k}
\end{align}
showing that the Chebyshev polynomials are orthogonal with respect to the weighted inner product defined by the left hand side of Eq.~\eqref{eq:inner}.

\subsection{Chebyshev Expansion of Complex-Valued Functions}\label{sub:Chebyshev Expansion of Complex-Valued Functions}
Chebyshev polynomials can be used to approximate a given complex-valued function $f$ via its Fourier series representation. To show that, we introduce the $2\pi$-periodic function\cite{Fox:1968up}
\begin{align}
	g(x)
	=f(\cos(x))
\end{align}
which can be represented in the interval (\emph{e.g.}, $-\pi<x<\pi$) in terms of its Fourier series as follows
\begin{align}
	g(x)
	=\sum_{k=0}^\infty (2-\delta_{k,0})a_k\cos(kx),\quad
	a_k
	=\frac{1}{\pi}\int_{0}^{\pi}g(x)\cos(kx)\,\mathrm{d}x.
\end{align}
Therefore
\begin{align}
	f(y)
	=g(\arccos(y))
\end{align}
can be represented in terms of the Chebyshev polynomials as follows
\begin{align}\label{eq:cheb_exp}
	f(y)
	=\sum_{k=0}^\infty \left(2-\delta_{k,0}\right)c_kT_k(y),
	\quad
	c_k
	=\frac{1}{\pi}\int_{-1}^1\frac{\mathrm{d}y}{\sqrt{1-y^2}}\,f(y)T_k(y),
\end{align}
for $y\in [-1,1]$. Equation~\eqref{eq:cheb_exp} is called the Chebyshev expansion of $f$ and it can be used to approximate $f$ as the linear combination of the first $N$ Chebyshev polynomials as follows
\begin{align}\label{eq:partial_Chebyshev}
	f(y)
	\approx S_Nf(y)
	=\sum_{k=0}^{N-1}\left(2-\delta_{k,0}\right)c_kT_k(y).
\end{align}
The coefficients $c_k$, defined by Eq.~(\ref{eq:cheb_exp}), are essentially the Fourier coefficients of the function $g$ that decay exponentially with $N$ for analytic functions\cite{Bergold:2020aa} (\emph{i.e.}, smooth in the complex domain) and thus provide fast convergence of the partial sums $S_Nf$. The resulting Chebyshev approximant $S_Nf$ is a polynomial of degree $N$, which is known to be close to the polynomial of the same degree with minimal error in the interval $[-1,1]$.\cite{Tal-Ezer:1989uj}

\subsection{Chebyshev Propagation in Discrete Representations}\label{sub:Chebyshev Propagation Method}
We obtain an approximation of the operator $\hat U=e^{-it\hat H}$ at time $t$ by considering the function $f(y)=e^{-ity}$ for which the coefficients $c_k$ defined according to Eq.~(\ref{eq:cheb_exp}) can be expressed in terms of the Bessel functions $J_k$ (of the first kind) as follows\cite{Abramowitz:1964aa}
\begin{align}
	c_k
	=(-i)^kJ_k(t)
\end{align}
yielding the following approximation
\begin{align}\label{eq:Cheby1}
	e^{-ity}
	\approx\sum_{k=0}^{N-1} \left(2-\delta_{k,0}\right)(-i)^kJ_k(t)T_k(y),
\end{align}
with $y\in[-1,1]$. Using a linear transformation of the argument $y$, we can restate Eq.~\eqref{eq:Cheby1} for an arbitrary Hermitian matrix $H\in\C^{D\times D}$ (where $D>1$ is a positive integer) with eigenvalues contained in a finite interval $[a,b]$ as follows
\begin{align}\label{eq:Cheby2}
	e^{-itH}
	\approx e^{-it^+}\sum_{k=0}^{N-1} \left(2-\delta_{k,0}\right)(-i)^kJ_k(t^-)T_k(H_0)
\end{align}
where we have introduced the rescaled variables $t^-,t^+\in\R$, and the matrix $H_0\in\C^{D\times D}$ with eigenvalues in $[-1,1]$ defined as follows
\begin{align}\label{eq:rescaling}
	t^\pm
	=\frac{t}{2}(b\pm a)
	\quad\text{and}\quad
	H_0
	=\frac{2}{b-a}\left(H-\frac{b+a}{2}\operatorname{I}_D\right)
\end{align}
where $\operatorname{I}_D$ is the $D\times D$ identity matrix.

Fast convergence is typically obtained for $e^{-ity}$ since it is a smooth function, although the number of required polynomials increases with $t$ since $e^{-ity}$ is oscillatory. Thus, a sufficiently large number $N$ of Chebyshev polynomials is needed to resolve the oscillations. In fact, it has been shown that the error falls like the $N$th order in $|t^-|/(2N)$ for sufficiently large $N$.\cite{Lubich:2008aa}

It is important to note that Eq.~\eqref{eq:Cheby2} can be used more generally than in the current implementation to approximate the solution to any linear system of the form $i\dot u=Hu$. Such linear systems typically arise in space discretization methods, including the Fourier collocation method, the Fourier Galerkin method, or the Hermite Galerkin method.\cite{Lubich:2008aa} So, we anticipate that the FTTC method should also be valuable for solving high-dimensional linear systems in a wide range of applications beyond the solution of the time-dependent Schr\"odinger equation.

\subsection{Discrete Tensor-Train Implementation}\label{sub:dtt}
Discrete tensor-train approximations of $e^{-it\hat H}$ are obtained by discretizing the $d$-dimensional space with a uniform grid of size $\Delta x_j>0$ for the $j$th nuclear coordinate, spanning the range $x_{j,\min}$ to $x_{j,\max}$, with $n_j>1$ points for each dimension $j=1,\dots,d$. Analogously, discrete tensor-train representations of wave functions are obtained as low-rank $d$-dimensional complex-valued tensor trains approximating
\begin{align}
	\tens{W}[k_1,\dots,k_d]
	=\Psi(x_k)
\end{align}
where $k=(k_1,\dots, k_d)$ are the indices of tensor-train entries corresponding to nuclear coordinate values $x_k=(x_{1,k_1},\dots,x_{d,k_d})$. The discrete tensor-train representation of $\tens{W}[k_1,\dots,k_d]$ is defined as follows
\begin{equation}\label{eq:tt}
	\tens{W}[k_{1},k_{2},\ldots,k_{d}]
	=\tens{W}_{1}[k_{1}]\tens{W}_{2}[k_{2}]\cdots\tens{W}_{d}[k_{d}],
	\quad1\le k_{j}\le n_j\textrm{ for all }j,
\end{equation}
where $\tens{W}_{j}[k_{j}]\in\reals^{r_{j-1}\times r_{j}}$ are matrices and $n_j$ are the number of grid points in the $j$th coordinate direction.

The action of the Hamiltonian $\hat H$ on a wave function $\Psi$ is represented by the Hermitian operator $\tens{\hat H}=\tens{\hat{T}}+\tens{V}$. The real-valued ``potential energy tensor''
\begin{align}
	\tens{V}[k_1,\dots,k_d]
	=V(x_{1,k_1},\dots,x_{d,k_d})
\end{align}
acts on $\tens{W}[k_1,\dots,k_d]$ as an element-wise multiplication operator (Hadamard product). The discrete kinetic energy operator
\begin{align}
	(\tens{\hat T}\tens{W})[k_1,\dots,k_d]
	\approx-\frac{1}{2m}\Delta_x\Psi(x_k),
\end{align}
is defined by the Laplacian $\Delta_x$ that acts as a multiplication operator in momentum space. Therefore, we apply the kinetic energy operator in momentum space by exploiting the highly efficient (linear scaling with dimensionality) implementation of multidimensional discrete Fourier transform of tensor trains to switch between position and momentum space. With the help of the fast Fourier transform (FFT), we therefore obtain a very efficient and accurate implementation of the discretized kinetic energy operator.

The discrete Hamiltonian is rescaled, according to Eq.~\eqref{eq:rescaling}, as follows
\begin{align}
	\tens{\hat H}_0
	=\frac{2}{E_{\max}-E_{\min}}\left(\tens{\hat H}-\frac{E_{\max}+E_{\min}}{2}\tens{\hat I}\right)
\end{align}
where $\tens{\hat I}$ denotes the identity on the tensor space. The bounds for the eigenvalues $E_{\min}$ and $E_{\max}$ depend on the extension of the grid and are given by
\begin{align}
	E_{\min}
	=\min_{k}\tens{V}[k_1,\dots,k_d],
	\quad
	E_{\max}
	=\max_{k}\tens{V}[k_1,\dots,k_d]+\frac{\pi^2}{2m}\left(\frac{1}{\Delta x_1^2}+\dots+\frac{1}{\Delta x_d^2}\right)
\end{align}
where we used $\Delta p_j=2\pi/(x_{j,\max}-x_{j,\min})$ for the grid size in momentum space of the $j$th coordinate, giving the maximum kinetic energy
\begin{align}
	\frac{1}{2m}p_{j,\max}^2
	=\frac{1}{2m}\frac{\pi^2}{\Delta x_j^2}.
\end{align}
Consequently, the solution $\Psi(t)$ is approximated with $N$ Chebyshev polynomials as follows
\begin{align}\label{eq:Cheby3}
	\Psi(t)
	=e^{-it\hat H}\Psi(0)
	\approx e^{-it^+}\sum_{k=0}^{N-1} \left(2-\delta_{k,0}\right)(-i)^kJ_k(t^-)T_k(\tens{\hat H}_0)\tens{W}_0
\end{align}
where $t^{\pm}=tE^{\pm}/2,\,E^{\pm}=E_{\max}\pm E_{\min}$ and $\tens{W}_0$ samples the initial wave function.

We implement Eq.~\eqref{eq:Cheby3} as a one-step propagator to compute $\Psi(t)$ directly from the initial data by using the Clenshaw algorithm\cite{Clenshaw:1955vn} (see Appendix~\ref{sec:Clenshaw}). Alternatively, we can obtain the time-dependent states $T_k(\tens{\hat H}_0)\tens{W}_0$ according to the recurrence relation Eq.~\eqref{eq:recurrence}
\begin{equation}
	\begin{split}
		T_0(\tens{\hat H}_0)\tens{W}_0
		&=\tens{W}_0,\\
		T_1(\tens{\hat H}_0)\tens{W}_0
		&=\tens{\hat H}_0\tens{W}_0,\\
		T_{k+1}(\tens{\hat H}_0)\tens{W}_0
		&=2\tens{\hat H}_0T_k(\tens{\hat H}_0)\tens{W}_0-T_{k-1}(\tens{\hat H}_0)\tens{W}_0,
		\quad\text{for $k\ge 1$}.
	\end{split}\label{eq:Tks2}
\end{equation}
The same Chebyshev propagation scheme described in this section for discrete tensor-train (TT) decompositions\cite{Oseledets.2011.2295} can be readily implemented using the continuous analogue functional tensor-train decomposition,\cite{Gorodetsky.2019.59} as described in the following section.

\section{Functional Tensor-Train Decomposition\label{sec:Function-Train}}

\subsection{Continuous Analogue of the Tensor-Train Decomposition}
Following refs 68 and 69, here we give a brief overview of the functional tensor-train (FT) format, which offers an efficient data compression scheme\cite{Gorodetsky.2019.59,Oseledets.2013.1}
\begin{equation}
	f(\x{1},\x{2},\ldots,\x{d})=\ftexp,\label{eq:ftlong}
\end{equation}
where $\uni{f}{k}{i}{j}:\xspace{k}\to\reals$, $\xspace{k}$ denotes the domain of the $k$th physical dimension, with $r_{0}=r_{d}=1$ for single-output functions such as polynomials or linear elements. The FT decomposition, introduced by Eq.~(\ref{eq:ftlong}), is a low-rank decomposition of multivariate functions in an analogous way that the TT decomposition,  Eq.~(\ref{eq:tt}), is a low-rank decomposition of multivariate arrays. A more compact expression for the FT analogue is obtained by viewing a function value as a set of products of matrix-valued functions,
\begin{equation}\label{eq:ft}
	f(\x{1},\x{2},\ldots,\x{d})=\mvf{F}_{1}(x_{1})\mvf{F}_{2}(x_{2})\ldots\mvf{F}_{d}(x_{d}),
\end{equation}
where each matrix-valued function $\mvf{F}_{k}:\xspace{k}\to\reals^{r_{k-1}\times r_{k}}$ is called a \textit{core} and can be visualized as an array of the univariate functions
\begin{equation}\label{eq:ftcore}
	\mvf{F}_{k}(x_{k})
	=\left[
	\begin{array}{ccc}\unif{k}{1}{1}(\x{k}) & \cdots & \unif{k}{1}{r_{k}}(\x{k})\\
		\vdots & \ddots & \vdots\\
		\unif{k}{r_{k-1}}{1}(\x{k}) & \cdots & \unif{k}{r_{k-1}}{r_{k}}(\x{k})
	\end{array}
	\right].
\end{equation}
If each univariate function is represented with $\pdimp{}{}{}$ parameters (for example, coefficients of a polynomial) and $r_{k}\le r$ for all $k$, then the storage complexity scales as $\mathcal{O}(dpr^{2})$. Comparing this representation with Eq.~\eqref{eq:tt}, we see a very close resemblance between the TT cores and the FT cores. Indeed, they are both matrices when indexed by a discrete index $i_{k}$ for the TT or a continuous index $x_{k}$ for the FT.

\subsection{Parameterizations of Low-Rank Functions}\label{sec:params}
The finer structure of FT format is described by the FT cores comprised of $\xdim$ sets of univariate functions $\left(\fcore{k}\right)_{k=1}^{\xdim}$. Each set could be different for different dimensions ({\em e.g.}, $2\pi$-periodic functions could represent physical dimensions corresponding to torsional angles, while Hermite polynomials could represent stretching modes). As a result, the full FT is parameterized through the parameterization of the set of univariate functions of each dimension, as chosen for the optimal representation of physical coordinates. Considering that $\pdimp{k}{i}{j}\in\integers_{+}$ denotes the number of parameters describing $\unif{k}{i}{j}$, and $\param\in\reals^{\pdim}$ the vector of parameters of all of the univariate functions, then there are a total of $\pdim\equiv\sum_{k=1}^{\xdim}\sum_{i=1}^{r_{k-1}}\sum_{j=1}^{r_{k}}\pdimp{k}{i}{j}$ parameters describing the FT representation.

The parameter vector $\param$ is indexed by a multi-index $\multi=(k,i,j,\ell)$ where $k=1,\ldots,\xdim$ corresponds to an input variable, $i=1,\ldots,r_{k-1}$ and $j=1,\ldots,r_{k}$ correspond to a univariate function within the $k$th core, and $\ell=1,\ldots,\pdimp{k}{i}{j}$ corresponds to a specific parameter within that univariate function. In other words, we adopt the convention that $\param_{\multi}=\param_{kij\ell}$ refers to the $\ell$th parameter of the univariate function in the $i$th row and $j$th column of the $k$th core.

The additional flexibility of the representation allows both linear and nonlinear parameterizations of univariate functions. In particular, the linear parameterization represents a univariate function as an expansion of basis functions $\left(\uni{\phi}{k\ell}{i}{j}:\xspace{k}\to\reals\right)_{\ell=1}^{\pdimp{k}{i}{j}}$ according to 
\begin{equation}\label{eq:linparam}
	\unif{k}{i}{j}(\x{k};\param)
	=\sum_{\ell=1}^{\pdimp{k}{i}{j}}\paramuni{k}{i}{j}{\ell}\uni{\phi}{k\ell}{i}{j}(\x{k}).
\end{equation}

\subsection{Low-Rank Functions vs Low-Rank Coefficients}\label{sec:lrf_vs_lrp}
For greater versatility, the FT can be used by independently parameterizing the univariate functions of each core, and both linear and nonlinear parameterizations are possible. As described below, the advantage of this representation includes a naturally sparse storage scheme for the cores.

An advantage of this type of structure is that it readily enables adaptivity when performing common multilinear algebraic operations with functions in low-rank format.\cite{Gorodetsky.2017.Continuous} For example, taking the product of two functions in low-rank format requires computing products between univariate functions in corresponding cores of the two functions. In particular, it requires computing the product between every combination of the functions. Because we store each univariate function separately, this product can accurately account for the complexity of the resulting univariate function. For instance, if the product of two third-order polynomials is considered, then a sixth-order polynomial will be stored. However, if a third-order polynomial is multiplied by a first-order polynomial, then only a fourth-order polynomial needs to be stored. In contrast, traditional tensor-based storage schemes would require storing each univariate function with the same number of polynomials. These advantages arise because we consider Chebfun-style continuous computation.\cite{driscoll2014chebfun} The salient point is that we consider univariate functions, rather than matrices or lower-order arrays, as the building blocks of low-rank representations. Another advantage is the availability of efficient computational algorithms for multilinear algebra that can adapt the representation of each univariate function individually as needed in the spirit of continuous computation pioneered by Chebfun.\cite{driscoll2014chebfun}

The TT/MPS format is a particular case of the general FT decomposition, naturally arising when two simplifying assumptions are made\cite{Mathelin.2014.243,Chevreuil.2014.897}
\begin{enumerate}
	\item linear parameterization of each $\unif{k}{i}{j}$;
	\item identical basis for the functions within each FT core, i.e., $\pdimp{k}{i}{j}=\pdimp{k}{}{}$ and $\uni{\phi}{k\ell}{i}{j}=\phi_{k\ell}$ for all $i=1,\ldots, r_{k-1}$, $j=1,\ldots,r_{k}$, and $\ell=1,\ldots,\pdimp{k}{}{}$;
\end{enumerate}
These assumptions transform the problem of storing low-rank functions to the problem of storing low-rank coefficients, allowing the use of discrete TT algorithms and theory. Both representations store the coefficients of a tensor-product basis $\phi_{k\ell}$ for all $k$ and $\ell$.

Function evaluations can be obtained from the coefficients defining the tensor $\tens{F}_{k}\in\reals^{r_{k-1}\times\pdimp{k}{}{}\times r_{k}}$ of the following form 
\begin{equation}\label{eq:ttcore}
	\tens{F}_{k}[:,\ell,:]
	=\left[
	\begin{array}{ccc}
		\paramuni{k}{1}{1}{\ell} & \cdots & \paramuni{k}{1}{r_{k}}{\ell}\\
		\vdots & \ddots & \vdots\\
		\paramuni{k}{r_{k-1}}{1}{\ell} & \cdots & \paramuni{k}{r_{k-1}}{r_{k}}{\ell}
	\end{array}
	\right],
\end{equation}
for $\ell=1,\ldots,\pdimp{k}{}{}$, by performing the following summation 
\begin{equation}\label{eq:low_rank_coeffs}
	f(x_{1},\ldots,x_{\xdim})
	=\sum_{\ell_{1}=1}^{\pdimp{1}{}{}}\cdots\sum_{\ell_{\xdim}=1}^{\pdimp{\xdim}{}{}}\tens{F}_{1}[:,\ell_{1},:]\cdots\tens{F}_{\xdim}[:,\ell_{\xdim},:]\phi_{1\ell_{1}}(x_{1})\cdots\phi_{\xdim\ell_{\xdim}}(x_{\xdim}).
\end{equation}
From Eqs.~\eqref{eq:ft},~\eqref{eq:ttcore}, and~\eqref{eq:low_rank_coeffs} we can see that the relationship between the TT cores $\tens{F}_{k}$ and the FT cores $\mvf{F}_{k}$ is 
\begin{equation}\label{eq:tt_to_ft}
	\mvf{F}_{k}(x_{k})
	=\sum_{\ell=1}^{\pdimp{k}{}{}}\tens{F}_{k}[:,\ell,:]\phi_{k\ell}(x_{k}),
\end{equation}
where the basis function multiplies every element of the tensor.
In other words, the FT cores represent a TT decomposition of the $p_{1}\times p_{2}\times\cdots\times p_{d}$ coefficient tensor of a tensor-product basis and inherit the properties of the TT decomposition.

\subsection{Operations in the FT Format}
Performing continuous multilinear algebra is one of the main advantages of the continuous framework. The operations of addition, multiplication, differentiation, integration, and inner products are easily performed for functions in the FT format as follows.\cite{Gorodetsky.2017.Continuous} Addition and multiplication of two functions are performed similarly to addition and multiplication of tensors in the TT format. For addition, the cores of $g(x)=f(x)+h(x)$ are 
\begin{align}
	\gcore{1}(x)
	=[\fcore{1}(x)\quad\hcore{1}(x)],\quad\gcore{k}(x)
	=\left[
	\begin{array}{cc}
		\fcore{k}(x) & \mat{0}\\
		\mat{0} & \hcore{k}(x)
	\end{array}
	\right],\quad\gcore{d}(x)
	=\left[
	\begin{array}{c}
		\fcore{d}(x)\\
		\hcore{d}(x)
	\end{array}
	\right],
\end{align}
for $k=2,\ldots, d$. For multiplication, $g(x)=f(x)h(x)$, we have
\begin{align}
	\gcore{k}(x)=\fcore{k}(x)\otimes\hcore{k}(x)\quad\text{ for }k=1,\ldots, d.
\end{align}
For both of these operations, the continuous functional decomposition has an important advantage compared to operations based on the discretized representation. Primarily, the advantage comes from the ability to add functions of differing discretization levels, e.g., functions represented with bases of different orders. In the discrete case, one can only add functions with identical discretizations.

The continuous nature of the FT also allows us to perform differentiation, as necessary to the implementation of the Laplacian, by differentiating scalar-valued functions that make up the corresponding core. For example, consider the partial derivative of a $d$-dimensional function $f$
\begin{align}
	\frac{\partial f}{\partial x_{k}}
	&=\mvf{F}_{1}\ldots\mvf{F}_{k-1}
	\left[
	\begin{array}{ccc}
		\ensuremath{\frac{d\ffiber{11}{k}}{dx_{k}}} & \cdots & \frac{d\ffiber{1r_{k}}{k}}{dx_{k}}\\
		\vdots & \ddots & \vdots\\
		\frac{d\ffiber{r_{k-1}1}{k}}{dx_{k}} & \cdots & \frac{d\ffiber{r_{k-1}r_{k}}{k}}{dx_{k}}
	\end{array}
	\right]\mvf{F}_{k+1}\ldots\mvf{F}_{d}.
\end{align}
When the univariate functions are expressed in, for example, a basis of orthonormal polynomials, then this operation is unique, well-defined, and computationally inexpensive.

Integration is widely used in Section~\ref{sec:Results} to compute expectation values. Integrating the multivariate functions scales linearly with dimensionality since it requires integrating over the one-dimensional functions in each core and then performing matrix-vector multiplication $d-1$ times as follows 
\begin{align}\label{eq:integrate}
	\int f(x)dx
	&=\int\fcore{1}(x_{1})\fcore{2}(x_{2})\ldots\fcore{d}(x_{d})dx_{1}\ldots dx_{d}\nonumber\\
 	&=\left(\int\fcore{1}(x_{1})dx_{1}\right)\left(\int\fcore{2}(x_{2})dx_{2}\right)\ldots\left(\int\fcore{d}(x_{d})dx_{d}\right)\nonumber\\
 	&=\mat{\Gamma_{1}}\mat{\Gamma_{2}}\ldots\mat{\Gamma_{d}}
\end{align}
where $\mat{\Gamma_{k}}=\int\fcore{k}(x_{k})dx_{k}$ contains entries $\mat{\Gamma_{k}}[i,j]=\int\ffiber{ij}{k}(x_{k})dx_{k}$ and the integral stands for an integral over an arbitrary domain. Furthermore, since each of the univariate functions is typically represented on a known basis, the integral is well defined, unique, and computationally inexpensive to obtain.

The inner product between two functions is another important operation essential for quantum dynamics simulations and computations of correlation functions. Naively, the inner product can be implemented by first computing the product $g(x)=f(x)h(x)$ and then integrating $g(x)$, requiring $\mathcal{O}(dr^{4})$ operations. However, this operation can be made more efficient by combining the operations needed for integration and multiplication. For example, Algorithm~\ref{alg:ftinner} uses an efficient computation of $v^{T}\left(\mat{A}\otimes\mat{B}\right)$ to perform the inner product in $\mathcal{O}(dr^{3})$ where $v$ is a vector and $\mat{A}$ and $\mat{B}$ are matrices.
\begin{algorithm}[H]
	\caption{$\texttt{ft-inner}$: Inner product between two functions in FT format\label{alg:ftinner}}
	\begin{algorithmic}[1]
		\REQUIRE Functions $f$ with ranks $r^{(f)}_k$ and $g$ with ranks $r^{(g)}_k$ in FT format
		\ENSURE $y = \int f(x)g(x) dx$
		\STATE $\bvec{y} = \int \gcore{1}(x_1) \otimes \fcore{1}(x_1) dx_1$
		\FOR {$k= 2 \ \TO \ d$}
			\STATE $\mat{Y} = \texttt{reshape}(\bvec{y}, r^{(f)}_{k-1}, r^{(g)}_{k-1})$
			\STATE $\mvf{T} = \fcore{k}(x)^T\mat{Y}$
			\STATE $\mvf{A} = \gcore{k}(x)^T \mvf{T}^T(x)$
			\STATE $\mat{Y} = \int \mvf{A}(x_k) dx_k$
			\STATE $\bvec{y} = \texttt{reshape}(\mat{Y}, 1, r^{(f)}_{k} r^{(g)}_{k} )$
		\ENDFOR
		\STATE $y = \bvec{y}[1]$
	\end{algorithmic}
\end{algorithm}
Furthermore, once in FT format, many other familiar operators may be applied to a function with relative ease. Consider the Laplacian $\Delta f(x)=g(x)=\sum_{k=1}^{d}\frac{\partial^{2}f(x)}{\partial x_{k}^{2}}$, necessary for implementation of the kinetic energy operator without having to rely on the Fourier transform. Written in this form, one can consider the Laplacian as the summation of $d$ functions $g_{k}(x)$ in function-train format, where
\begin{equation}
	g_{k}(x)
	=\frac{\partial^{2}f(x)}{\partial x_{k}^{2}}.
\end{equation}
The second derivative is implemented core-by-core in the space of univariate functions. The second derivatives of univariate functions are computed only once, which exploits the benefits of the continuous representation and avoids the need for explicit calculation of the Fourier transform required by grid-based methods.

\section{Functional Tensor-Train Chebyshev Propagation}\label{sec:TTCheb}
Wavepackets and operators are efficiently represented in terms of low-rank functional tensor trains (FT) or discrete tensor trains (TT). The decompositions are constructed analytically or interpolated with the cross approximation as implemented in the Compressed Continuous Computation ($C^{3}$) library\cite{Gorodetsky.2016.Compressed} in terms of linear element expansions or the TT-Toolbox.\cite{Oseledets.2020.TT} Operations are computed in the position-space representation, including the kinetic energy operator in the FT representation, which is computed analytically from the Laplacian. In the discrete TT representation, the kinetic energy operator is computed numerically in momentum space. Algebraic manipulations are followed by rounding schemes to avoid an artificial growth of the rank.

The functional tensor-train algebra discussed in Section~\ref{sec:Function-Train} is then employed to express the individual Chebyshev polynomials of the Hamiltonian as applied to the initial wavepacket, as discussed in Section~\ref{sec:ChebApproach}. The codes are available in public domain.\cite{codes} The minimal and maximal potential energy surface values required for rescaling the Hamiltonian in the Chebyshev scheme are determined either analytically or through constrained nonlinear optimization to avoid calculation of the multidimensional potential energy surface at all position-space grid points considered. Individual Chebyshev polynomials are determined as either tensor trains or function trains via the recurrence relation Eq.~\eqref{eq:recurrence} or the action of the propagator on the wavefunction is determined directly from the Clenshaw algorithm, see Appendix~\ref{sec:Clenshaw}. The resultant dynamics is analyzed via calculation of survival amplitudes and wavepackets.

\section{Chemical Model\label{sec:Model}}
\begin{figure}[H]
	\centering
	\hspace{2.5cm}
	\includegraphics[width=0.7\textwidth]{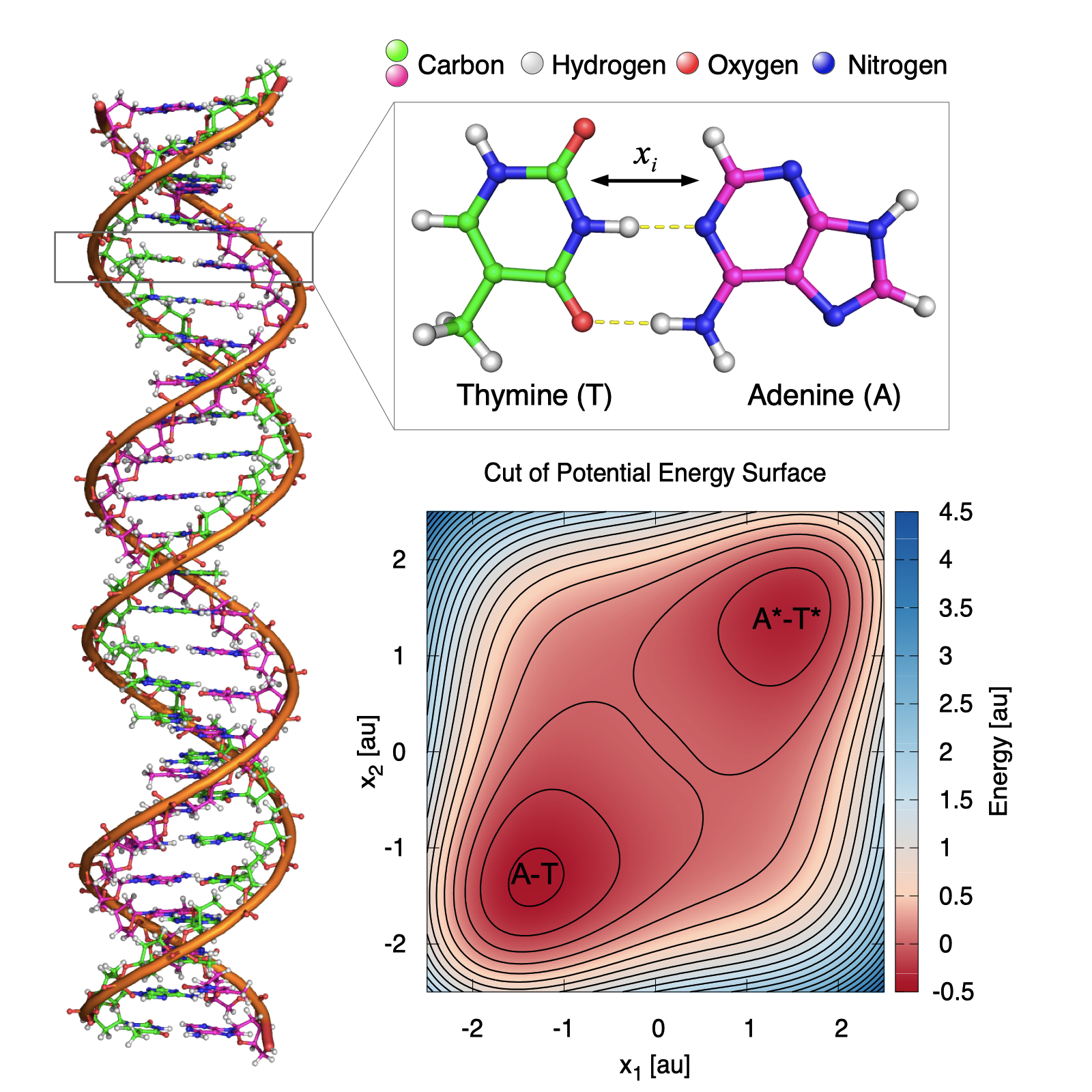}
	\caption{(Left) DNA strand of adenine-thymine base pairs (top right) with a two-dimensional slice of the model potential energy surface Eq.~\eqref{eq:potential} (bottom right).	\label{fig:PES}}
\end{figure}

We simulate the dynamics of protons in a 50-dimensional model of hydrogen-bonded DNA adenine-thymine base pairs, described by the model potential energy surface\cite{Godbeer.2015.13034,Soley.2021.3280}
\begin{equation}\label{eq:potential}
	V\left(x_{1},x_{2},\ldots,x_{d}\right)
	=\sum_{i=1}^{D}\alpha\left(0.429~x_{i}-1.126~x_{i}^{2}-0.143~x_{i}^{3}+0.563~x_{i}^{4}\right)+\sum_{i>1}^{D}\alpha\beta\left(x_{i}x_{i-1}\right)
\end{equation}
where $\alpha=0.1\text{ au}$ determines the energy scaling of the model potential and $\beta$ is the hydrogen-bond coupling parameter (see Fig.~\ref{fig:PES}). Each $x_i$ describes the coordinate of proton motion in an individual adenine-thymine (A-T) pair as it tautomerizes from the energetically favored amino-keto A-T form to the isomeric imino-enol A*-T* form. The coupling term parameterized by $\beta$ provides a model of interaction between base pairs.
The resulting 50-dimensional model potential involves strongly anharmonic modes, which are beyond the reach of the grid-based Chebyshev approach or other quantum dynamic methods based on full-grid representations. The molecular system also provides a challenging test case for low-rank tensor-train-based dynamics, as the potential energy surface becomes increasingly demanding as the coupling parameter is dialed up to $\beta=-2\text{ au}$. Therefore, the resulting wavepackets can reach maximal ranks of over $r_\text{max}=100$ without truncation in the FT representation.

We examine the ability of the Chebyshev method to simulate isomerization processes by considering the initial state, introduced by Eq.~\eqref{eq:InitialState}, that represents the excited A*-T* tautomer with width $\alpha=1\text{ au}$, position $x_{0,i}=1\text{ au}$, momentum $p_{0,i}=0\text{ au}$, and mass $m=1\text{ au}$. A position-space vector of grid length $L=10\text{ a.u.}$ in the TT format and a position-space region of $x\in[-5,5]\text{ au}$ in the FT format (with $N_{x}=N_{p}=2^{5}$ equal divisions in position space) is used to capture the full extent of the reactive coordinate oscillation between the two isomers. The wavepacket is computed at intermediate times (with a time step of $\tau=0.01\text{ au}$) by defining each intermediate time as an endpoint. A basis set of $N_{\text{poly}}=50$ polynomial terms is used in the Chebyshev expansion to accurately represent the dynamics in both the TT and FT formats for the DNA system.

\section{Results}\label{sec:Results}
\begin{figure}[H]
	\centering
	\hspace{2.5cm}
	\includegraphics[width=1\textwidth]{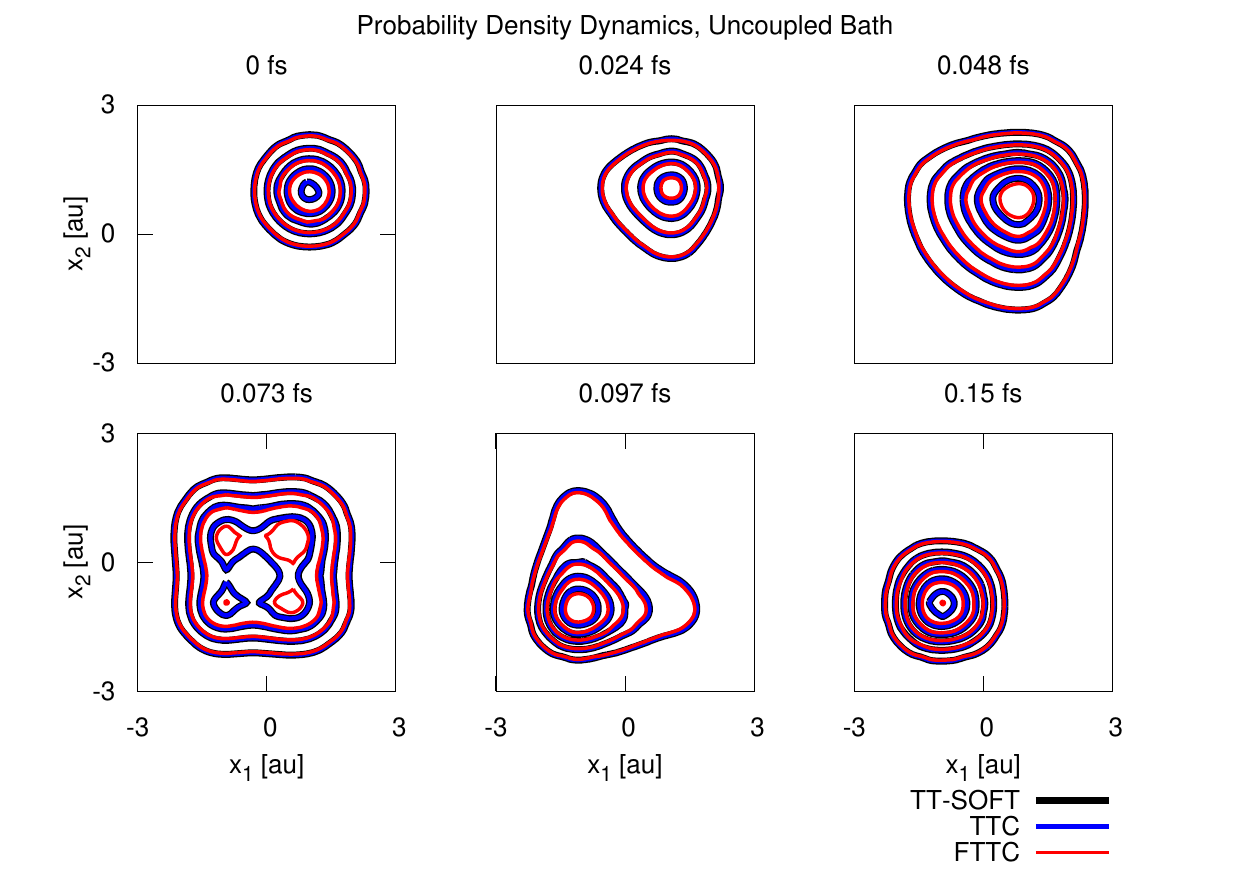}
	\caption{Comparison of two-dimensional slices of the 50-dimensional time-dependent wavepacket obtained from FTTC simulations (red line) and its discrete TT implementation (blue line) as compared to benchmark TT-SOFT (black) simulations of tautomerization quantum dynamics for uncoupled ($\beta=0\text{ au}$) DNA base pairs.\label{fig:50UncoupledDensity}}
\end{figure}

\begin{figure}[H]
	\centering
	\hspace{2.5cm}
	\includegraphics[width=1\textwidth]{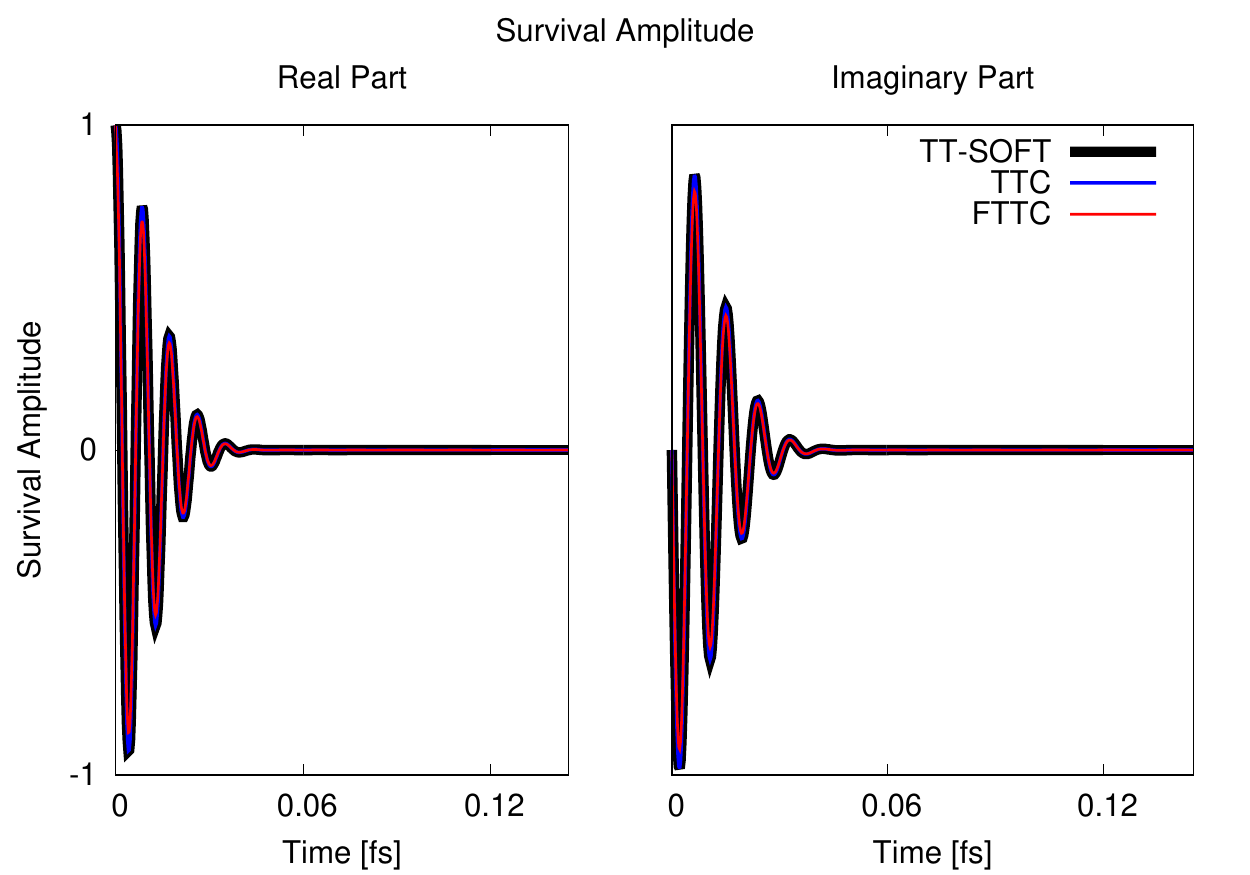}
	\caption{Comparison of survival amplitudes from simulations of the time-dependent wavepacket evolving on a 50-dimensional potential energy surface for the tautomerization dynamics of uncoupled ($\beta=0\text{ au}$) DNA base pairs, including the real (left) and imaginary (right) parts, obtained with FTTC (red line) and its discrete TT implementation (blue line) as compared to benchmark TT-SOFT (black).\label{fig:50UncoupledAutocorrelation}}
\end{figure}

\begin{figure}[H]
	\begin{center}
		\includegraphics[width=1\textwidth]{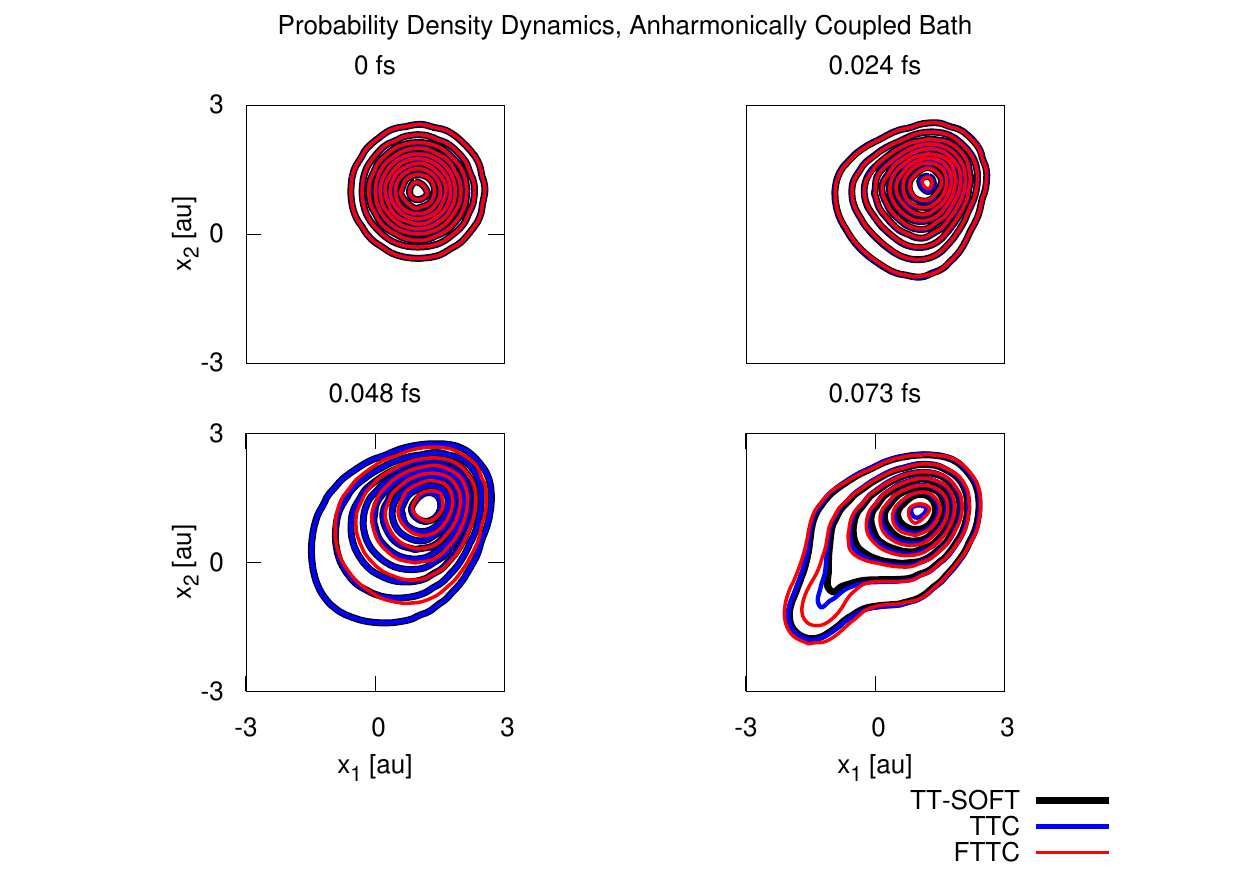}
	\end{center}
	\caption{Comparison of two-dimensional slices of the 50-dimensional time-dependent wavepacket obtained from FTTC simulations (red line) and its discrete TT implementation (blue line) as compared to benchmark TT-SOFT (black) simulations of tautomerization quantum dynamics for coupled ($\beta=-2\text{ au}$) DNA base pairs.\label{fig:50CoupledDensity}}
\end{figure}

\begin{figure}[H]
	\begin{center}
		\includegraphics[width=1\textwidth]{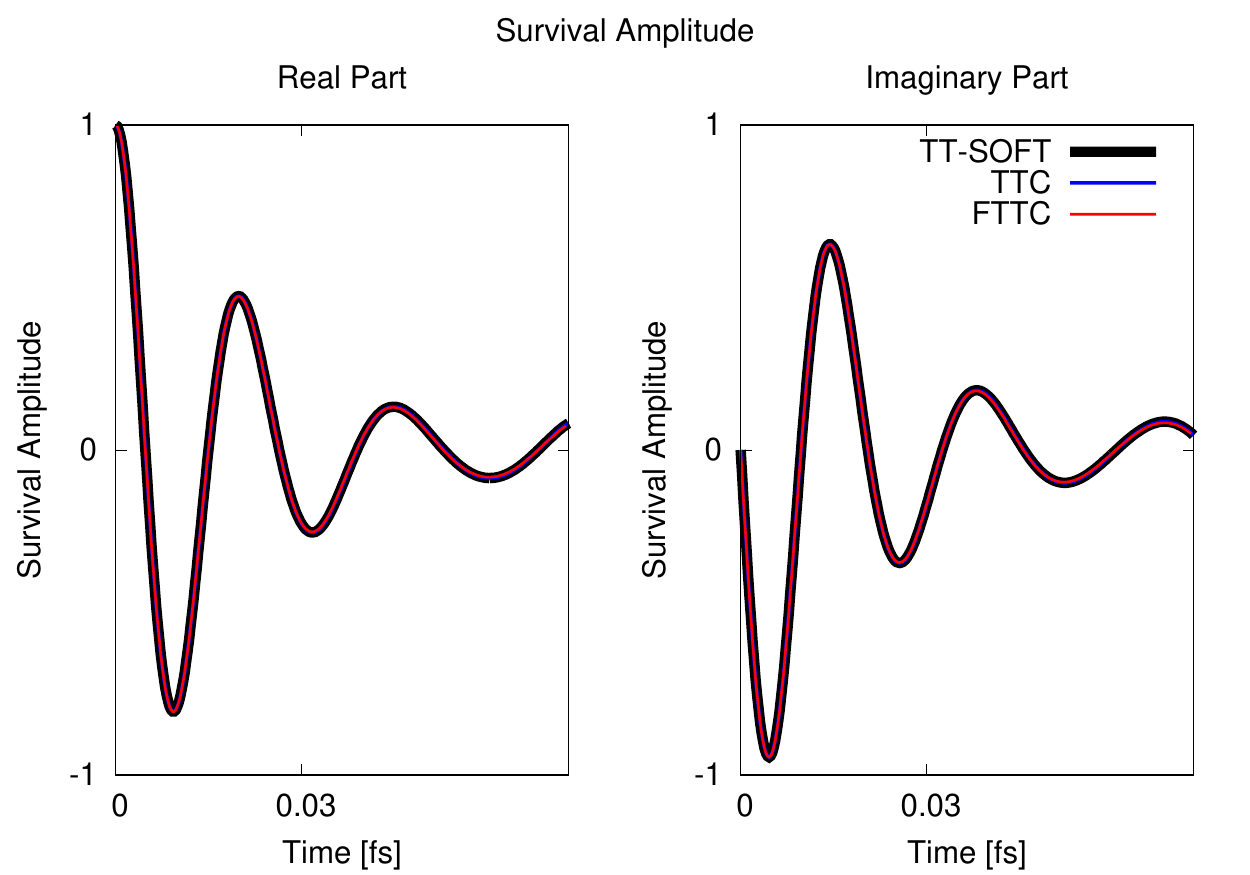}
	\end{center}
	\caption{Comparison of survival amplitudes from simulations of the time-dependent wavepacket evolving on a 50-dimensional potential energy surface for the tautomerization dynamics of coupled ($\beta=-2\text{ au}$) DNA base pairs, including the real (left) and imaginary (right) parts, obtained with FTTC (red line), its discrete TT implementation (blue line), and benchmark TT-SOFT (black).\label{fig:50CoupledAutocorrelation}}
\end{figure}

Figures~\ref{fig:50UncoupledDensity} and~\ref{fig:50UncoupledAutocorrelation} show benchmark calculations of FTTC simulations for the 50-dimensional tautomerization of uncoupled DNA base pairs, as compared to the discrete TT implementation and TT-SOFT simulations. The corresponding simulations for coupled DNA base pairs $\beta=-2\text{ au}$ are compared in Figs.~\ref{fig:50CoupledDensity} and~\ref{fig:50CoupledAutocorrelation}. Comparison slices of the time-dependent wavepacket along two of the 50 dimensions and survival amplitudes show excellent agreement between the methodologies and efficient performance even without relying on high-performance computing facilities.
The simulations are initialized by a Gaussian,
\begin{equation}\label{eq:InitialState}
	\Psi_{0}\left(x\right)
	=\prod_{i=1}^D\sqrt[4]{\frac{\alpha}{\pi}}\exp\left(-\frac{\alpha}{2}\left(x_i-x_{0,i}\right)^{2}+\text{i}p_{0,i}\left(x_i-x_{0,i}\right)\right),
\end{equation}
with $x_{0,i}=1\text{ au}$ and $p_{0,i}=0\text{ au}$, corresponding to a displaced tautomeric form along the double-well potential energy surface characterizing the energy change as a function of the proton displacement. The resulting dynamics leads to the motion of the wavepacket into the well of the energetically favored tautomer as the isomerization due to proton dynamics proceeds in the 50-dimensional space of the model system.
%

\section{Discussion}
Numerically exact quantum dynamical methods that rely on full-grid representations are not applicable to high-dimensional model systems since they require computational resources that scale exponentially with dimensionality. Even the standard grid-based implementation of the Chebyshev method, renowned for its ability to achieve accuracy within machine precision, has been limited in applications to nuclear quantum dynamics to model systems with no more than four atoms. Here, we have shown how to extend the capabilities of the Chebyshev propagation scheme to high-dimensional systems in terms of the FTTC algorithm. We anticipate that the resulting FTTC methodology will be useful not only for simulations of quantum reaction dynamics in general but also as a general method to obtain numerical solutions of linear systems in high dimensionality, typically arising from space discretization in many other types of applications. Furthermore, the functional train decomposition should also find wide applicability in studies requiring computations of gradients, integrals, and correlation functions of systems with high dimensionality.

With regards to the basis functions, we note that the functional tensor-train representation can implement suitable choices of univariate basis functions that could be ideal for data compression in chemistry, for example, waveforms or Gaussian functions, which are common to both wavepacket propagation methods and electronic structure calculations alike. In general, representations that require $O(n^d)$ data points in a $d$-dimensional grid with $n$ points for each direction would require at most $O(dnr^2)$ data points for a maximum rank $r$ in a discrete tensor-train representation and only $O(dpr^2)$ data points in functional tensor-train format where $p$ is the number of parameters, which represents a significant reduction in computational cost ideal for modeling molecular systems.

\section{Acknowledgments}
M.~B.~S.~acknowledges financial support from the Yale Quantum Institute Postdoctoral Fellowship.
V.~S.~B.~acknowledges support from the NSF Grant no. CHE-1900160 and high-performance computing time from NERSC and the Yale High-Performance Computing Center. A.~A.~G. was supported by the AFOSR Computational Mathematics Program under the Young Investigator Program.

\section*{Appendix}
\appendix
\section{Clenshaw Algorithm}\label{sec:Clenshaw}
The direct computation of the Chebyshev expansion Eq.~\eqref{eq:Cheby3} based on the usual summation algorithm has two disadvantages: (1) all summands have to be kept in the memory of the computer, which can be very expensive in practical applications since the tensors $T_k(\tens{\hat H}_0)\tens{W}_0$ (and also their low-rank approximations) are typically large objects, and (2) it is known that the worst-case error generated by the floating point operations grows proportionally to the number $N$ of summands.\cite{Higham:2002ta}. We therefore use the Clenshaw algorithm,\cite{Clenshaw:1955vn} which offers a stable alternative to evaluate linear combinations of polynomials that satisfy a linear recurrence relation such as the Chebyshev polynomials.\cite{Fox:1968up}\\

Assuming that for given coefficients $c_0,c_1,\dots,c_{N-1}\in\C$ we are interested in the value of the partial Chebyshev sum Eq.~\eqref{eq:partial_Chebyshev}, the Clenshaw algorithm replaces the summation by the evaluation of the following backward recurrence system
\begin{align}
	\begin{cases}
		B_r(y)=2yB_{r+1}(y)-B_{r+2}(y)+c_r,\quad r=N-1,\dots,0;\\
		B_N(y)=0,\quad B_{N+1}(y)=0;
	\end{cases}
\end{align}
and then expresses the partial Chebyshev sum as
\begin{align}
	\sum_{k=0}^{N-1}(2-\delta_{k,0})c_kT_k(y)
	=B_0(y)-B_2(y).
\end{align}
To obtain the approximation of the wavefunction $\Psi(t)$, we adapted the Clenshaw algorithm by first solving the backward recurrence system
\begin{align}
	\begin{cases}
		\tens{B}_r=2\tens{\hat H}_0\tens{B}_{r+1}-\tens{B}_{r+2}+(-i)^rJ_r(t^-)\tens{W}_0,\quad r=N-1,\dots,0;\\
		\tens{B}_N=0,\quad\tens{B}_{N-1}=0;
	\end{cases}
\end{align}
and then computing the approximant
\begin{align}
	\Psi(t)
	\approx e^{-it^+}\left(\tens{B}_0-\tens{B}_2\right).
\end{align}
We note that this numerically stable procedure needs to keep only three tensors in memory.

\section{TOC Graphic}
\includegraphics[width=1.\textwidth]{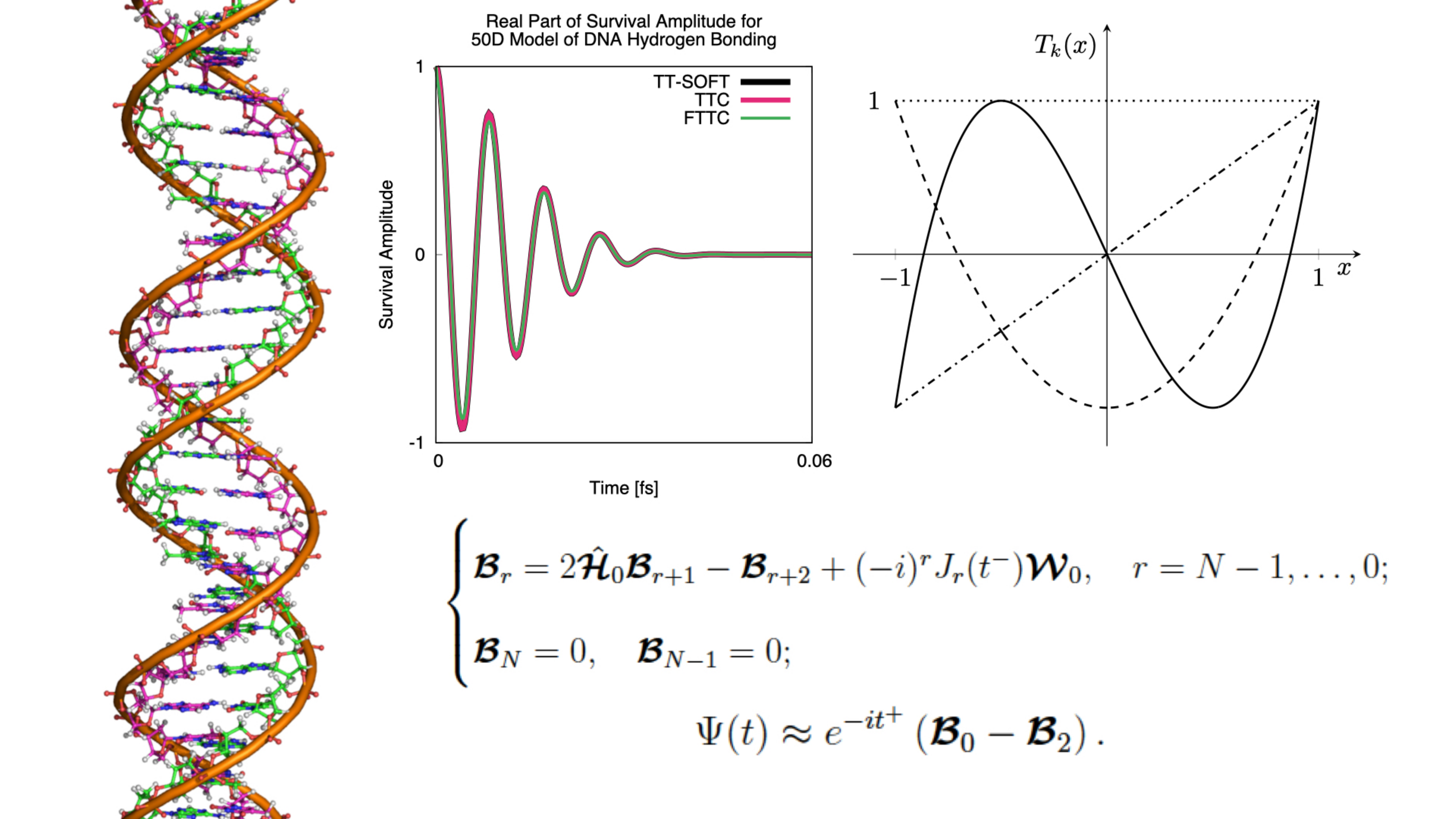}

\bibliographystyle{plain}
\bibliography{fttc}

\end{document}


\newpage

\section{Computational Performance}\label{sec:ComputationalPerformance}
The favorable computational performance of the tensor-train Chebyshev (TTC) method is demonstrated through analysis of a system whose time evolution is known exactly, a Gaussian coherent state in the harmonic oscillator potential
\begin{equation}
	V(x)
	=\sum_{i=1}^D \frac{1}{2}m\omega^2x_i^2
\end{equation}
of dimension $D=2$, mass $m=1\,\text{au}$, and frequency $\omega=1\,\text{au}$.

The benchmark analytic time evolution of the Gaussian state is determined recognizing that a coherent state $\left|\alpha\right\rangle $ can be expanded in terms of the harmonic oscillator eigenstates states $\left|n\right\rangle $ of energy $E_n=(n+1/2)\hbar\omega$ as follows
\begin{equation}
	\left|\alpha\right\rangle
	=e^{-\frac{1}{2}\left|\alpha\right|^{2}}\sum_{n=0}^{\infty}\frac{\alpha^{n}}{\sqrt{n!}}\left|n\right\rangle
\end{equation}
where $\alpha$ is a complex number that indicates the displacement of the coherent state from the origin in phase space, such that the coherent state is an eigenstate of the annihilation operator
\begin{align}
	\hat{a}\left|\alpha\right\rangle
	=\alpha\left|\alpha\right\rangle
\end{align}
where
\begin{equation}
	\hat{a}\left|n\right\rangle
	=\sqrt{n}\left|n-1\right\rangle
\end{equation}
and an eigenstate of the time-evolved annihilation operator in the Heisenberg picture
\begin{align}
	\hat{a}\left(t\right)\left|\alpha\right\rangle
	&=e^{\frac{i}{\hbar}\hat{H}t}\hat{a}e^{-\frac{i}{\hbar}\hat{H}t}\left|\alpha\right\rangle\\
	&=\alpha e^{-\frac{1}{2}\left|\alpha\right|^{2}}\sum_{n=1}^{\infty}\frac{\alpha^{n-1}}{\sqrt{n-1!}}e^{-\frac{i}{\hbar}\left(E_{n}-E_{n-1}\right)t}\left|n-1\right\rangle\\
	&=\alpha e^{-i\omega t}e^{-\frac{1}{2}\left|\alpha\right|^{2}}\sum_{n=0}^{\infty}\frac{\alpha^{n}}{\sqrt{n!}}\left|n\right\rangle\\
	&=\alpha e^{-i\omega t}\left|\alpha\right\rangle
\end{align}
The time evolution of the coherent state in the Schr\"odinger picture is then determined by the action of the annihilation operator
\begin{equation}
	\hat{a}\left|\alpha\left(t\right)\right\rangle
	=\alpha(t)\left|\alpha\left(t\right)\right\rangle
\end{equation}
with the time-dependent displacement of the coherent state
\begin{equation}
	\alpha\left(t\right)
	=e^{-i\omega t}\alpha\left(0\right)
\end{equation}
which yields the position-space representation of the time-evolved coherent state as a coherent state of the form
\begin{equation}
	\Psi\left(x,t\right)
	=\langle x|\alpha\rangle
	=\left(\frac{\omega}{\pi\hbar}\right)^{1/4}\exp\left(-\frac{1}{2}\left|\alpha\left(0\right)\right|^{2}+\frac{\omega}{2\hbar}x^{2}-\left(\sqrt{\frac{\omega}{\hbar}}x-\frac{e^{-i\omega t}\alpha\left(0\right)}{\sqrt{2}}\right)^{2}\right)
\end{equation}

\begin{figure}[H]
	\centering
	\includegraphics[width=0.75\textwidth]{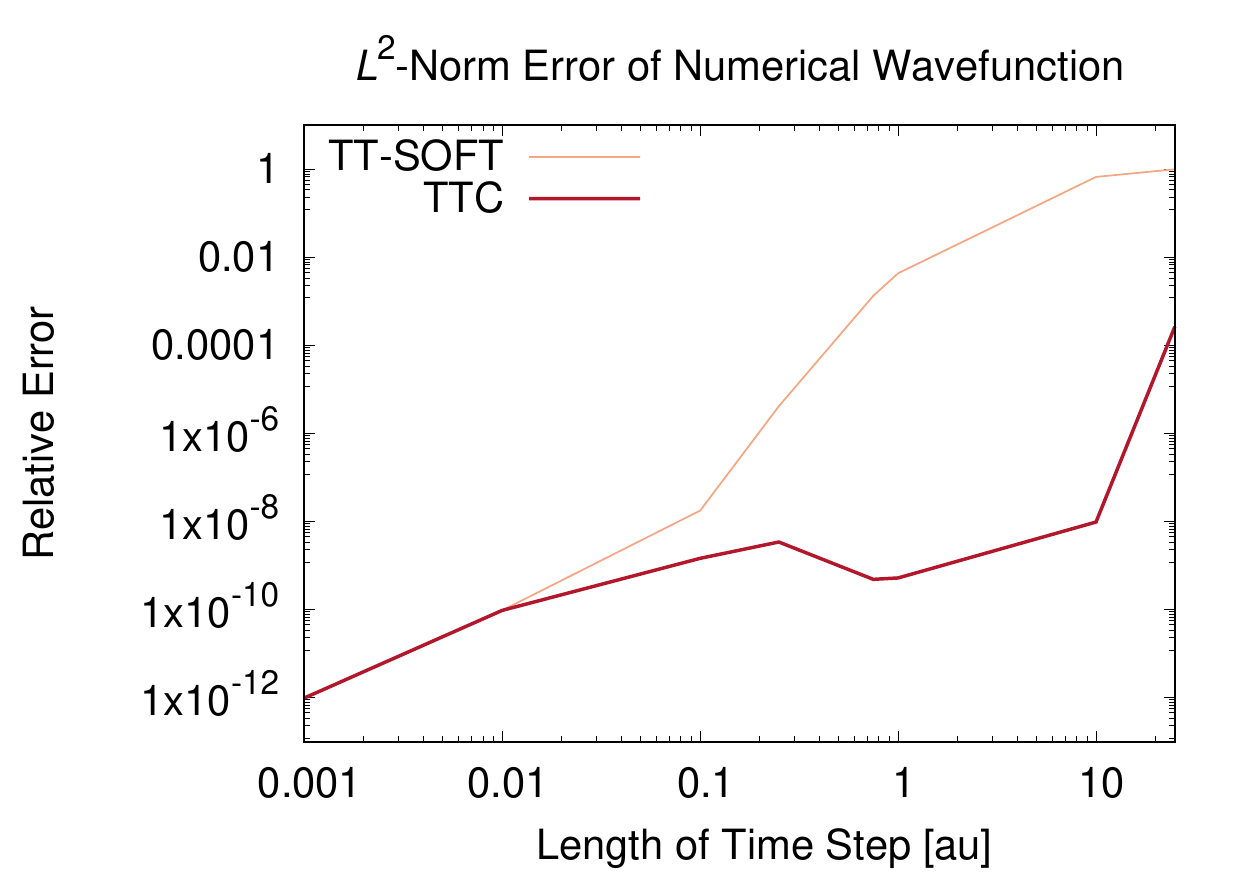}
	\caption{The $L^2$-norm error of the numerical wavefunction is significantly lower for TTC (red line, $N=750$ Chebyshev polynomials) than TT-SOFT (orange line) for long time steps.\label{fig:L2SOFTComparison}}
\end{figure}

\begin{figure}[H]
	\centering
	\includegraphics[width=0.75\textwidth]{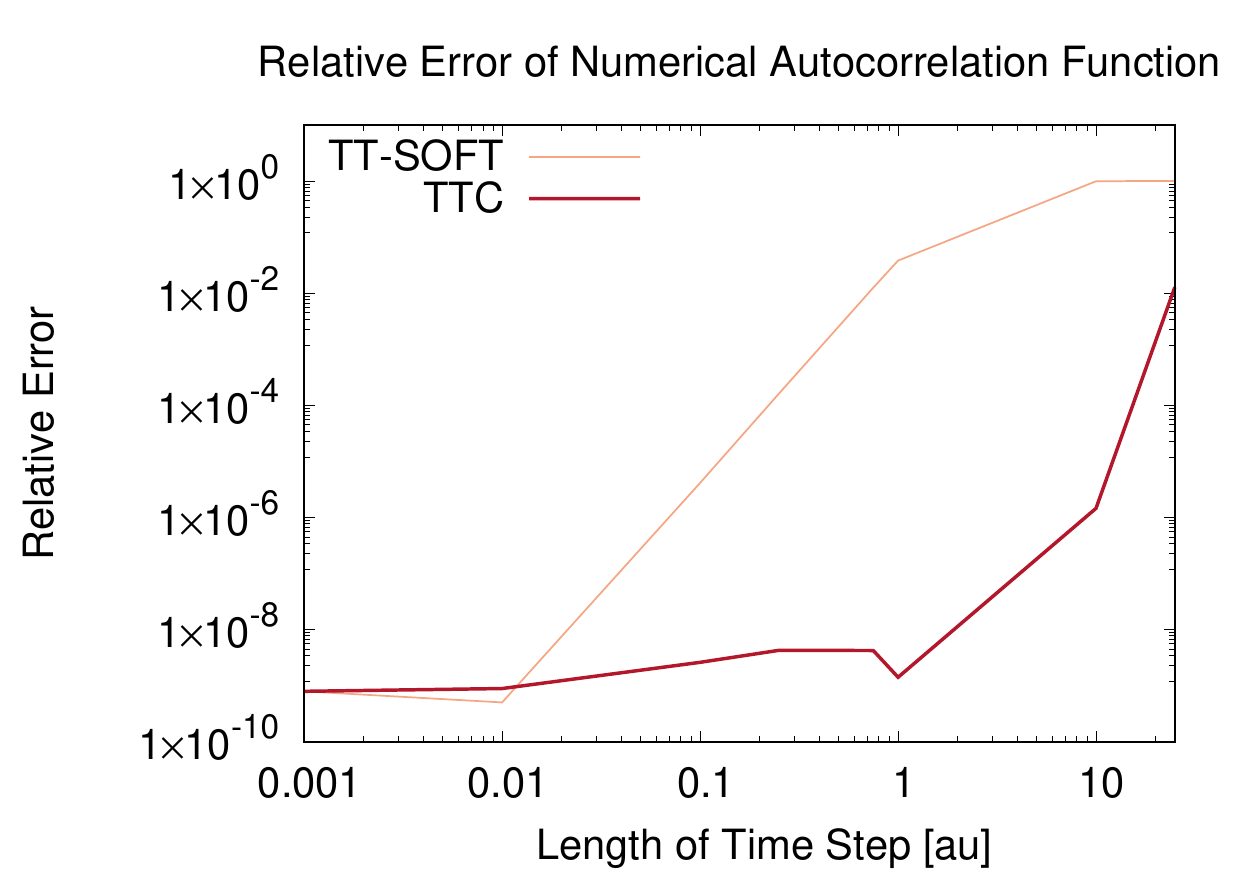}
	\caption{TTC (red line, $N=750$ Chebyshev polynomials) significantly reduces the relative error associated with determination of the autocorrelation function of the benchmark system relative to the state-of-the-art short-time TT-SOFT method (orange line) for long time steps.\label{fig:AutocorrelationSOFTComparison}}
\end{figure}

\begin{figure}[H]
	\centering
	\includegraphics[width=0.75\textwidth]{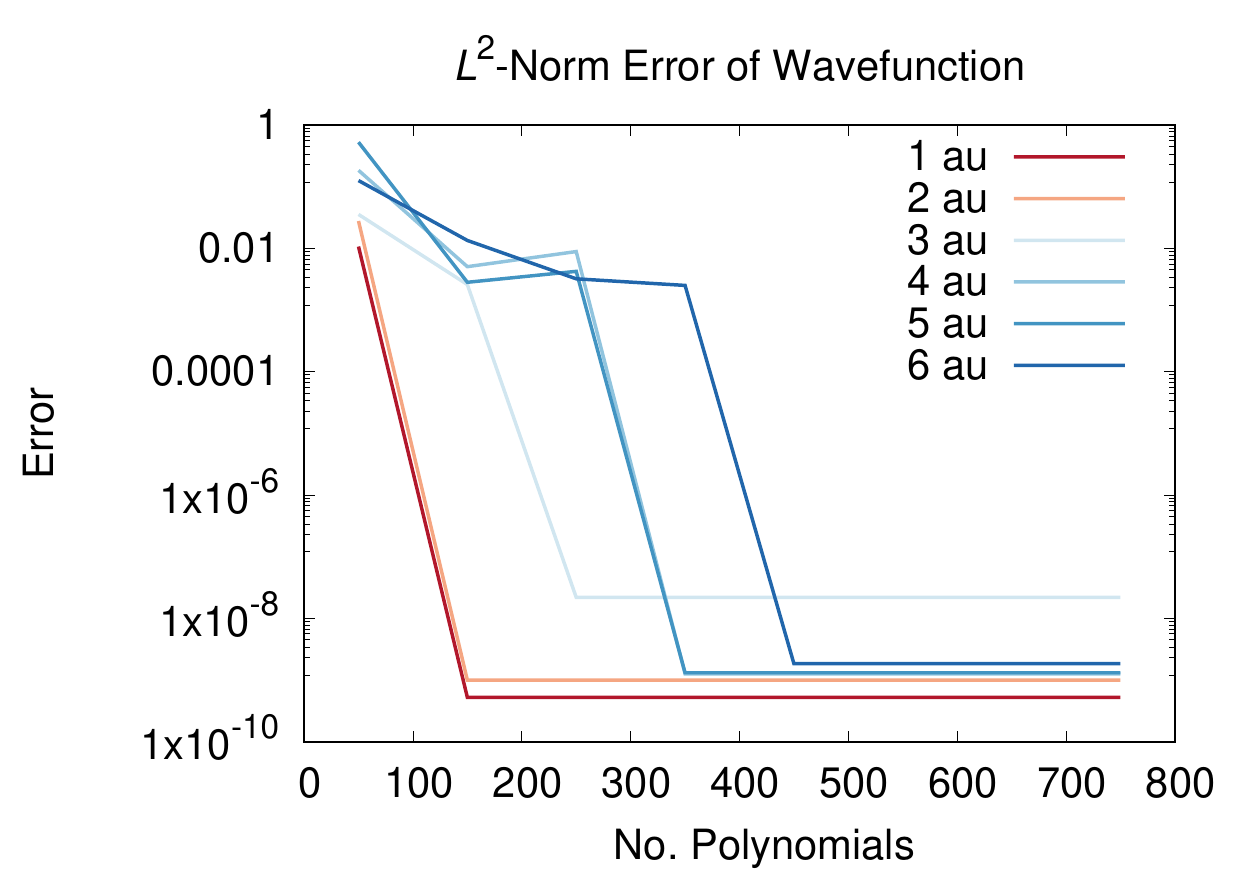}
	\caption{$L^2$-norm error of the wavefunction in TTC as a function of the number of Chebyshev polynomials for varying final times.\label{fig:L2ErrorPolynomials}}
\end{figure}

We find the wavepacket determined by tensor-train Chebyshev (TTC) quantum dynamics is significantly more accurate than the short-time Tensor-Train Split Operator Fourier Transform (TT-SOFT) approach for long time steps, as shown in Fig.~\ref{fig:L2SOFTComparison}. Whereas the error of the $L^2$-norm of the wavefunction is nearly equivalent for both methods for short time steps, TTC produces the wavepacket with several orders of magnitude lower $L^2$-norm error for time steps on the order of 100 to 1000 times longer than those required by TT-SOFT given a sufficient number of Chebyshev polynomials in the expansion of the propagator. This suggests TTC can require fewer costly Fourier transform computations than TT-SOFT for long time steps.
Likewise, TTC molecular dynamics is found to be more accurate relative to TT-SOFT for computation of the autocorrelation function over long time steps (see Fig.~\ref{fig:AutocorrelationSOFTComparison}). For short time steps, TTC and TT-SOFT closely agree, and TTC reduces the relative error of the autocorrelation function by orders of magnitude at the longest time steps considered. We find TTC yields the autocorrelation function with only one percent error with a time step 100 times longer than the maximum time step that can be used to accurately simulate the dynamics with TT-SOFT. The TTC molecular dynamics method therefore successfully maintains accuracy and, where desired, avoids calculation of the wavefunction at intermediate time steps for calculation of the autocorrelation function at the final time.
Examination of the relative error of the $L^2$-norm of the wavefunction as a function of the number of terms in the Chebyshev expansion confirms the ability of the TTC algorithm to achieve high accuracy solutions for long time steps given a sufficient number of Chebyshev polynomials, as shown in Fig.~\ref{fig:L2ErrorPolynomials}. The TTC approach requires fewer than 200 Chebyshev polynomials to accurately determine the propagator for a final time of $1\,\text{au}$, such that the method is efficient for short-time propagation, which is beneficial for computation of the wavefunction at intermediate time steps. As expected, the number of polynomials required for accurate simulation of the benchmark system's dynamics increases for direct propagation of the initial wavepacket for larger final times. The expansion is shown to converge for final times up to $6\,\text{au}$ for fewer than $500$ polynomials, which demonstrates the robustness of TTC for simulation of long-time dynamics in agreement with the measure of Chebyshev accuracy previously observed in the one-dimensional case.
